 \documentstyle[epsf,epsfig,rotate]{aa}

\newcommand{\be}{\begin{equation}}
\newcommand{\ee}{\end{equation}}

\newcommand{\ls}{\raisebox{-.8ex}{$\buildrel{\textstyle<}\over\sim$}}

\newcommand{\anrev}{{\it ARA\&A, }}
\newcommand{\apj}{{\it ApJ, }}

\newcommand{\mnr}{{\it MNRAS, }}

\newcommand{\pasp}{{\it PASP }}
\newcommand{\ana}{{\it A\&A, }}
\newcommand{\anas}{{\it A\&AS, }}

\newcounter{pp3}
\addtocounter{pp3}{3}

\begin{document}
 \thesaurus{02(02.01.2; 02.13.2; 08.09.2 AA Tau; 08.13.1; 08.16.5)}

\title
{ On disc driven inward migration of resonantly coupled planets
with application to the system around  GJ876 }

\author{ M.D. Snellgrove,  J.C.B. Papaloizou \and  R.P. Nelson}

\institute{ Astronomy Unit, 
 Queen Mary, University of London, Mile End
 Rd, London E1 4NS} 
 
\offprints{m.d.snellgrove@qmw.ac.uk}

\date{Received /Accepted}

\def\LaTeX{L\kern-.36em\raise.3ex\hbox{a}\kern-.15em
         T\kern-.1667em\lower.7ex\hbox{E}\kern-.125emX}



\titlerunning{Resonance coupling of planets in a disc}
\authorrunning{Snellgrove, Papaloizou \& Nelson}

\maketitle

\begin{abstract}
We consider two protoplanets gravitationally interacting with each
other and a protoplanetary disc. The two  planets orbit
interior to a tidally maintained disc cavity  while the disc interaction induces
inward migration.   When the migration is slow enough, the more rapidly
migrating outer protoplanet  approaches and becomes
locked in a $2:1$ commensurability  with the inner one. This is maintained
in subsequent evolution. 
We study this evolution using a simple analytic model,
full  hydrodynamic 2D simulations of the disc planet system
and  longer time  N body integrations incorporating simple 
prescriptions for the effects of the disc on the planet orbits.
The eccentricities of the protoplanets are  found to be determined
by the migration rate and  circularization rate 
induced in the outer planet orbit by the external disc.

We apply our results to the recently discovered resonant planets
around GJ876.  Simulation shows that a disc with parameters expected for
protoplanetary discs causes trapping in the  $2:1$ commensurability when
the planets orbit in an inner cavity  and 
that eccentricities  in the observed range
may be obtained.

\end{abstract}


\section{Introduction} \label{intro}

The recent discovery of  extrasolar giant planets orbiting around
nearby solar--type stars (Marcy \& Butler 1998, 2000) has stimulated renewed interest in the
theory of planet formation. 
The objects observed so far have masses, $M_p$, that
are characteristic of giant planets

\noindent ($0.4 \; {\rm M}_J \; \ls \; M_p \ \ls \; 11 \; {\rm M}_J$), 
$M_J$ denoting a Jupiter mass.
The orbital semi-major axes are  in the range 

\noindent $0.04 \; {\rm AU} \; \ls \; a \; \ls \; 2.5 \; {\rm AU}$, and orbital 
eccentricities in the range $ 0.0 \; \ls \; e \; \ls \; 0.67$ 
(Marcy \& Butler 2000).  

Disc-protoplanet interactions have been invoked to explain the presence
of giant planets orbiting close to their host stars  through inward 
orbital migration
induced  through disc-protoplanet tidal interaction ( eg.
Papaloizou, Terquem \& Nelson 1999,  Lin et al 2000).
Up to now, for the most part extrasolar planets appear to be isolated.
However, a few multiple systems are known. The configuration of these
may contain important information about their origin and possible migration history.
Of special interest is the recently discovered system around GJ876.
This is found to be close to a $2:1$ commensurability.     
Such a configuration is indeed suggestive of orbital migration.
Commensurable satellite systems such as the Galilean satellites are 
thought to  owe their origin to migration induced by tidal interaction
with the central planet ( eg. Goldreich 1965).

Recent simulations of single  protoplanets in the  observed
mass range (Kley 1999, Bryden et al. 1999, Lubow, Seibert \& Artymowicz 1999)
interacting with a disc with parameters thought to be typical
of protoplanetary discs, but constrained to
be in circular orbit, indicate gap formation and
upper mass limit consistent with the observations. 
Nelson et al. (2000)  
relaxed the assumption of fixed circular orbits, 
found inward migration  and  that the
disc-protoplanet interaction leads to strong eccentricity damping.
Due to accretion onto the central star,
an inner cavity was formed in the disc  interior to which 
the protoplanet orbited.

Simulations of two planets interacting with a disc have been performed
by Kley (2000), Bryden et al (2000), and  Masset \& Snellgrove (2001).
So far inward migration of two planets locked into a $2:1$ commensurability
has not been simulated. However, Bryden et al (2000) found a tendency
for gap material between the two planets in fixed circular
orbits  to be cleared, ending up interior to the inner
planet orbit or exterior to the outer planet orbit.

Taken together the above results suggest a natural outcome 
of two protoplanets interacting with a disc is that they orbit interior
to an inner disc cavity while the external disc causes inward migration
of the outer orbit. This catches up the inner orbit
leading to the possibilty of resonant interaction.

It is the purpose of this paper to investigate  such resonant interaction
and whether, for reasonable
protoplanetary disc models,
it leads to a locking such that the planets subsequently
migrate maintaining  the commensurability.
Similar  behaviour  occurs as a result  of the tidally induced migration
of the Gallilean satellites
(eg. Goldreich 1965, Lin \& Papaloizou 1979).

Here we shall assume the prior evolution of the system
leads to the orbital separation of the planets being slightly
larger than that required for a strict $2:1$ commensurability
without considering the history in detail as 
it is beyond the scope of this paper. However, we comment
that this might have been complicated with the planet masses
varying with time through mass accretion from the disc.

Analytic methods, N body integrations and direct  two dimensional  numerical simulations
are used to investigate the evolution and found to give consistent results. 

We consider two protoplanets gravitationally interacting with each
other and a protoplanetary disc. The two  planets orbit
interior to a tidally maintained disc cavity  while the disc interaction induces
inward migration.  This migration is of type II
and so is regulated by the magnitude of the disc viscosity.
When the migration is slow enough, the more rapidly
migrating outer protoplanet  approaches and becomes
locked in a $2:1$ mean motion
commensurability  with the inner one. The commensurability persists 
in subsequent evolution. The eccentricities of the protoplanets are increased 
by the resonant perturbations and damped by circularization which occurs
through interaction with the disc (eg. Goldrech \& Tremaine 1981).
A balance is achieved in which the protoplanet eccentricities are
determined 
by the migration rate and  circularization rate 
induced in the outer planet orbit by the external disc.
 
We apply our results to the recently discovered resonant planets
around GJ876.  Simulation shows that migration
induced by  a disc with parameters expected for
protoplanetary discs  results in 
trapping in the  $2:1$ commensurability when
the planets orbit in an inner cavity. Eccentricities in the observed range
may be obtained.
Further studies using $N$ body integrations indicate that the planetary 
system will remain stable for at least $2 \times 10^{7}$ orbits
when the external disc is removed.

In section (\ref{mod}) we describe a simple analytic model of two migrating
protoplanets in a $2:1$ commensurability. The eccentricities of the protoplanet orbits
are related to the migration rate, and  circularization rate induced by the disc.
In section (\ref{simulation}) we describe a simulation of two planets orbiting
in an inner disc cavity. Parameters appropriate to GJ867 are adopted.
This demonstrates resonant trapping and that eccentricities of the observed magnitude
may be produced.  

In section (\ref{Nb}) we describe N body calculations confirming the above conclusions
and indicating the long term stability of the system.
Finally in section~(\ref{blah}) we discuss our results.

\section{ A simple model } \label{mod}
We  consider a system consisting
of $2$  planets and a primary star
moving under  their gravitational attraction.
When there are no disc interactions
and  the motion is conservative the system is conveniently expressed
in Hamiltonian form using Jacobi coordinates (eg. Sinclair 1975).
The  coordinates, ${\bf r}_2,$  of inner planet of reduced  mass $m_2$
are referred to the central star of mass $M_*$
and the coordinates of the  outer planet, ${\bf r }_1,$
of reduced  mass $m_1$ are referred
to the  centre of mass of the central star and inner planet.
The Hamiltonian can be written  correct to second order in the planetary masses as
\begin{eqnarray} H & = &  {1\over 2} ( m_1 | \dot {\bf r}_1|^2 +m_2| \dot {\bf r}_2|^2)
- {GM_{*1}m_1\over  | {\bf r}_1|} - {GM_{*2}m_2\over  | {\bf r}_2|} \nonumber \\
& - &{Gm_{1}m_2\over  | {\bf r}_{12}|}
 +  {Gm_{1}m_2 {\bf r}_1\cdot {\bf r}_2
\over  | {\bf r}_{1}|^3}.
\end{eqnarray}
Here $M_{*1}=M_*+m_1, M_{*2}= M_* + m_2 $ and
$ {\bf r}_{12}= {\bf r}_{2}- {\bf r}_{1}.$
The Hamiltonian can be expressed in terms of the  osculating  semi-major axes, eccentricities
and longtitudes of periastron $a_i,e_i,\varpi_i, i=1,2$ respectively   as well as the
longtitudes $\lambda_i,$  and the time
$t.$ We recall that $\lambda_i = n_i(t-t_{0i}) + \varpi_i,$ with $n_i$ being the mean
motion and $t_{0i}$ giving the time of periastron passage.
The energy is given by  $E_i = -Gm_iM_{*i}/(2a_i),$
and the angular momentum $h_i = m_i\sqrt{GM_{*i}a_i(1-e_i^2)}$  which may  be
used  to describe the motion  instead of $a_i$ and $e_i.$
 
Only terms first order in the eccentricities and involving the
resonant angles $\phi = 2\lambda_1-\lambda_2-\varpi_1, $ and
$\psi = 2\lambda_1-\lambda_2-\varpi_2,$  are retained  in the expansion (see Sinclair 1975).
Then the perturbing  part of the Hamiltonian $(\propto m_1 m_2)$ can be written
\be H _{12}= -{Gm_1m_2\over a_1}\left({Be_1\cos\phi} -{Ce_2\cos\psi}\right), \ee
where $C=2b_{1/2}^{(2)}({\alpha}) +(1/2)\alpha d b_{1/2}^{(2)}({\alpha})/d\alpha$ and 
$B=1.5b_{1/2}^{(1)}({\alpha}) +(1/2)\alpha d b_{1/2}^{(1)}({\alpha})/d\alpha -2\alpha.$
Here $b_{1/2}^{(n)}$ denotes the usual Laplace coefficient 
and $\alpha= a_2/a_1.$ From now on we replace $M_{*i}$ by $M_*.$
\subsection{Orbital precession}
We may also  take into account additional gravitational forces that
may produce precession of the planetary orbits. These could result from
the
mass distribution provided by the  disc exterior to the outer planet 
or the time averaged mass distribution of the planets
themselves. These effects are  not included in the resonant Hamiltonian
given above.  Accordingly we add to it another Hamiltonian
\be H_{int} = - {1\over 2}m_1 n_1 a_1^2 \omega_{pr1}e_1^2
 - {1\over 2}m_2 n_2 a_2^2 \omega_{pr2}e_2^2 . \label{Nonaxi}\ee
Here, as can be verified from the equations of motion, we have  adopted a
parameterization such that the precession frequency induced in the orbit
of $m_i$ is  $\omega_{pri}(a_i).$ This prescription does not
allow the precession frequecy to depend on other quantities.
However, appropriate matching can be carried out for
a particular case under consideration.
 
\subsection{Basic Equations}
The equations of motion are derived from: \\
${dE_i}/{dt} =-n_i\partial H' / \partial \lambda_i -(n_1 T +D)\delta_{i1}, 
$ \\
${dh_i}/{dt}  =-{\partial H'  / \partial \lambda_i} -{\partial H'  / \partial \varpi_i} -T\delta_{i1},$ \\ 
$ {d\lambda_i  / dt} = n_i +n_i {\partial H' / \partial E_i} + {\partial H'  / \partial h_i}, $ \\
${d\varpi_i / dt} ={\partial H'  / \partial h_i}, $ \\
with $H' = H_{12} + H_{int}.$ 
These can be obtained from
Hamilton's equations (eg. Brouwer \& Clemence 1961) to which we have added, for the outer planet $m_1,$
an additional external torque $- T$
with  an associated orbital energy  loss rate $n_1 T$  together with  additional
orbital energy  dissipation rate $D.$ The torque and dissipation rate  could be produced by
tidal interaction with the disc leading to inward migration and orbital circularization.
 
We thus obtain to  lowest order in the  planetary eccentricities and perturbing masses.
\begin{eqnarray} {dn_1\over dt} & = & {6n_1^2 m_2 \over M_*  }\left(Be_1\sin\phi -Ce_2\sin\psi\right) \nonumber \\
& + &{3 n_1 a_1\over G M_*m_1}\left(n_1 T +D\right) \label{first}
\end{eqnarray}
\be {dn_2\over dt} =-{3n_2^2 m_1 a_2\over M_* a_1 }\left(Be_1\sin\phi -Ce_2\sin\psi\right)\ee
\be {de_1\over dt} =-{ m_2 n_1\over M_* }B\sin\phi
-{D a_1\over GM_* m_1 e_1}\ee
\be {de_2\over dt} ={ m_1 n_2 a_2\over M_* a_1 }C\sin\psi \ee
\be {d\phi \over dt} = 2n_1- n_2 -
{m_2\over e_1 M_* }n_1 B\cos\phi
-\omega_{pr1}   \label{last2}\ee
\be {d\psi \over dt} =  2n_1- n_2 +
{ m_1 a_2\over  e_2 a_1 M_* }n_2 C\cos\psi -\omega_{pr2}
\label{last}\ee
\subsection{Stationary solutions}
When no migration or circularization  occurs ($T=D=0$)
equilibrium solutions may  exist
such that $\psi $ and
$\phi$ are either zero or $\pi.$   Each of $n_1,n_2,e_1,e_2$ are then constant.
 
A relation between the eccentricities  then follows from   (\ref{last2}) and (\ref{last})
in the form
\begin{eqnarray}
& &e_2a_1( e_1  M_* \omega_{pr1}+
 m_2 n_1 B \cos \phi)  = \nonumber \\ & & \;\;\;\;  e_1 (-m_1 a_2 n_2 C\cos\psi+
  e_2 a_1 M_* \omega_{pr2})  \label{eccp}.\end{eqnarray}
This condition matches the precession rates of the orbits
of the two planets.
Also $2n_1=n_2.$
Noting that the eccentricities are
positive, when they are of very small magnitude, the precessional terms
become negligible and there is a solution with $\psi=0, \phi =\pi$
or $\psi=0, \phi =\pi.$ In either case
we have
\be
 m_2 a_1  e_2 n_1 B  =  m_1 a_2 e_1n_2 C \label{eccpop}.\ee
For larger eccentricities the precessional terms may become important in
(\ref{eccp}) and then solutions with $\psi=0, \phi =0,$ may occur.
Then (\ref{eccp}) gives 
\be
 m_2 a_1  e_2 n_1 B + m_1 a_2 e_1n_2 C=  
 e_1 e_2 a_1 M_* ( \omega_{pr2} -\omega_{pr1}).  \label{eccpr}\ee

For stable solutions, when perturbed,  the angles may
undergo librations about their equilibrium
points (eg. Sinclair 1975). There are two frequencies
of oscillation $\nu_1,\nu_2$  being given for any $\psi, \phi $
 in the limit of
small eccentricities by $\nu_1^2 = (m_2 n_1 B)^2/(M_*e_1)^2,$
and $\nu_2^2 =(m_1n_2a_2C)^2/(a_1M_* e_2)^2.$
We look for a solutions with migration which are
 close to   stable solutions
of this type.
 
\subsection{Resonant Migration}
We look for solutions of (\ref{first}-\ref{last}) corresponding
to the situation where the two planets migrate inwards locked
in resonance with $n_1/n_2$  maintained nearly equal to $1/2$ while the
eccentricities remain nearly constant. The tendency for
the resonant coupling to excite the eccentricities is counterbalanced
by circularization through the action of
$D \equiv (GM_* m_1 e_1^2) /(a_1 t_c)$ which defines the circularization
time for $e_1.$ Similarly (\ref{first}) defines an inward  migration timescale
$t_{mig}= GM_* m_1/(3 T a_1n_1).$
 
We begin by  supposing  that the angle $\psi$ executes a
libration about zero such that the mean rate of change of $e_2$ is zero.
Similarly the mean rates of change of $n_1$ and $n_2$ induced by $\psi$ are zero.
Such a libration is seen
in simulations. We also suppose the angle $\phi$ either librates  or
circulates but in such a way that the correspondingly  induced mean rates of change
are not zero. The simplest example is when the angle executes a  very
small or even zero  amplitude  libration
about  a value slightly offset from zero or $\pi$ (eg. Lin \& Papaloizou 1979).
 
We suppose the circulation/libration  periods to be short compared
to the timescale of migration
so that averaging is possible.
We denote the average of $n_1 e_1\sin\varphi $ by $\delta.$
Then averaging (\ref{first}-\ref{last}) gives
\be {\overline{{1\over n_1}{dn_1\over dt}}}={6 m_2 B\delta\over M_* }
+{3 a_1 \over G M_*m_1}\left(n_1 T +D\right) \label{first1}\ee
\be {\overline{{1\over n_2}{dn_2\over dt}}}=-{6 m_1 B a_2 \delta\over M_* 
a_1}\ee
\be {\overline {{de_1^2\over dt}}}=-{2 m_2  B \delta\over M_* }
-{2D a_1 \over GM_* m_1 } \equiv 0 \label{eca}\ee
\be {\overline{{d\phi \over dt}}} =
-{m_2\over e_1 M_* }n_1 B\cos\phi-{ m_1 a_2 \over e_2 M_* a_1 }n_2 C
-\omega_{pr1} +\omega_{pr2}.
\label{last1}\ee
Here on the left hand sides
the overline denotes the time average and for simplicity
of notation this has been dropped from the right hand sides.
The resonance condition
also implies
${\overline{{1\over n_2}{dn_2\over dt} }}=
{\overline{{1\over n_1}{dn_1\over dt} }}.$
Using this and eliminating $\delta$ from the above, we obtain
for the rate of increase of $n_1$ through  migration
\be {\overline{{1\over n_1}{dn_1\over dt} }}
={3 a_1\over G M_*m_1}\left(n_1 T +D\right)
\left({m_1a_2\over m_1a_2+m_2a_1}\right).\ee
Also (\ref{eca}) gives for the eccentricity balance
\be e_1^2 = {t_c m_2a_1\over 3 t_{mig}(2 m_1 a_2 + m_2a_1)}. \label{e1} \ee
 
The above  determines the eccentricity of the outer planet
$e_1$ as a function of $t_c$ and $t_{mig}.$
For a system with $m_1/m_2 =3,$ we get
$e_1 \sim \sqrt{0.07t_c/t_{mig}}.$

\noindent For $e_1$ in the $0.01$ range we need $t_c \sim10 $~orbits
if $t_{mig} \sim 10^4$~orbits.
 
The eccentricity of the inner planet is determined
by  equation(\ref{last1}). For small amplitude librations
this
is given by  (\ref{eccp}).  That would still apply when $\phi$
is circulating provided  the cosines are time averaged
and the mean circulation rate is small.

\section{A simulation of two migrating resonantly 
coupled planets}\label{simulation}

The protoplanetary disc is numerically simulated using an  
Eulerian 2D hydrodynamic code. 
 The code used is
a modified version of NIRVANA, which has been described, tested and used
successfully elsewhere on a similar problem
involving interacting planets  (Masset \& Snellgrove \cite{Masset1}). 
 Incorparated with 
the hydrodynamic code is a 4th-order Runge-Kutta  integrator which is
used to 
 evolve the orbits of the
 two planets. The  gravitational  forces calculated
from the disk model are  used  in the equations of motion
 of the 
planets, and the disc itself responds to the planetary potential. Hence 
the system  evolves in a self-consistent fashion.  
In order to obtain the 
long integration times needed for  simulations
of this type  the FARGO algorithm 
(Masset \cite{Masset}) is applied.
However, tests have shown that the results are not affected
by this.

 We use a 2D cylindrical $(r,\varphi)$  grid with 200  
radial zones distributed uniformly
 between $r=0.4$ and $r=3.47$ in dimensionless units
 and 300 azimuthal zones. We 
 apply outflow conditions
 at the inner boundary to simulate the accretion of disc 
material onto the central star.

\subsection{Physical setup}

We attempt to simulate the resonance locking of the system GJ876 via 
tidally induced migration of the planets, using 
plausible values of the disc parameters. The disc 
is assumed to be thin and isothermal, with constant aspect ratio $h/r = 
0.07$, and a constant Shakura \& Sunyaev (\cite{Shakura})
 $\alpha$-viscosity prescription  with ${\alpha} = 2 \times 
10^{-3}$ is 
adopted. The two planets are initially in circular orbits coplanar with 
the disc, at radial locations $r_{1} =1.0$ and $r_{2} =0.6.$ Hence 
the outer planet is located outside the exact $2:1$ commensurability
($r_{2} =0.63$). The 
planet masses are chosen to
correspond to the minimum mass ratios 
obtained from 
observations (Marcy et al. \cite{Marcy4}). With masses normalised so that 
stellar 
mass is $M_{*} = 1$, this corresponds to $m_{1} = 6 \times 10^{-3}$ and 
$m_{2} 
= 1.8 \times 10^{-3}$. The planet masses are fixed as the planets are 
assumed to 
be no longer accreting material from the disc.

The disc is prescribed an initial surface density $\Sigma_{0}$ 
corresponding to 
what  would  give a disc
mass of $2 \times 10^{-3}$ within the orbit of the outer planet. However 
we
assume that both planets are located inside a tidally truncated cavity 
located at
$r < 1.3$, with  low surface density $\Sigma_{\mathrm{cavity}} = 
0.01\Sigma_{0}$. 
This cavity  is supposed to have already 
been cleared by the tidal action of the two planets. Between $1.3 < r < 
1.5$  the surface density 
is  prescribed such that $\ln \Sigma$   linearly joins to $\Sigma_0.$.

\subsection{Results}

\begin{figure}
\begin{center}
\epsfig{file=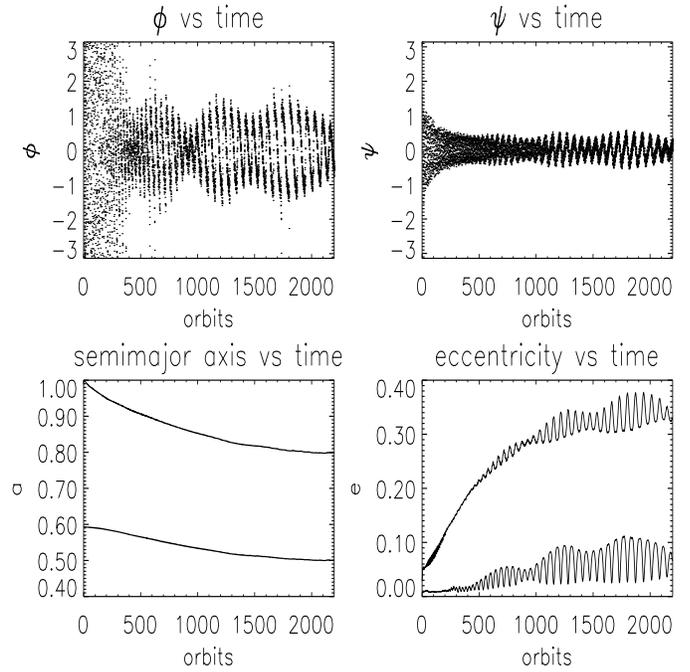, height=9cm, width=9cm}
\end{center}
\caption{The  lower plots show the semi major axes $a$ (lower left) 
and eccentricities $e$ (lower right) for both planets. The  upper plots 
show 
the resonance angles $\phi$ and $\psi$. The time unit is measured in 
orbit  periods of the disc material at $r=1.$}
\label{Fig1}
 \end{figure}

\begin{figure}
\begin{center}
\epsfig{file=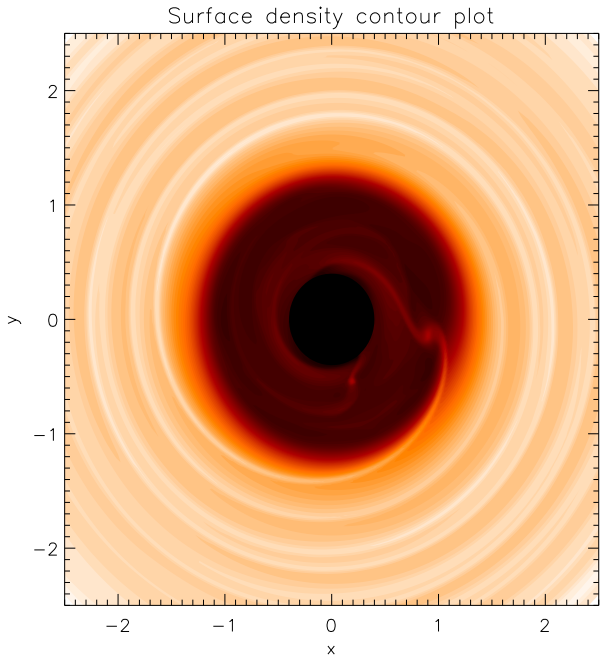, height=9cm, width=9cm}
\end{center}
\caption{Surface density plot for inner regions ofthe disc at a time 
corresponding to 300 orbits. Darker areas on the plot correspond to 
regions of lower surface density.
The disc cavity is clearly visible, as are the density waves 
outside the cavity. Inside the cavity are the wakes of the two planets.} 
\label{Fig2}
\end{figure}

The tidal interaction of the planets with the disc material causes the 
planets to migrate inwards (see Figure \ref{Fig1}). The inner planet is 
deep 
within the cavity and only interacts with  low surface density material
 and thus 
migrates slowly. The 
outer planet has its outer $2:1$ lindblad resonance located outside the 
cavity and in the body of the disc.
 Hence there is more material exerting a negative torque on 
the planet, and 
therefore it migrates faster, despite its larger mass. The ratio 
of semi-major axes $a_{1} / a_{2}$ of the planets decreases until the 
planets 'lock' into a $2:1$ commensurability with $n_{2} \approx 2n_{1}$ 
at a 
time 
 $t \approx 400$ orbits. Both  planets then subsequently migrate inwards a 
further 10\%  maintaining this ratio, showing the resonance to be robust.

Figure \ref{Fig1}
 also shows the calculated values of the resonance angles $\phi$ 
and 
$\psi$. Once the commensurability lock has occured, these are both 
librating about zero. The resonant interaction causes a eccentricity growth of both 
planets, the growth halting at around average values of $e_{1} = 0.06$ and 
$e_{2} = 0.34$ although both eccentricities exhibit variations around 
these average values. The 
perihelion angles of the two planets oscillate around the alignment 
position, being the natural stable state.

The disc cavity remains at low density, and the cavity edge diffuses 
slowly in on the viscous diffusion timescale, following the outer planet. 
Figure \ref{Fig2}  shows a surface density plot of the inner regions 
of the disc.

\subsection{Relation to analytic model and observations}

The simulation confirms that tidally induced migration of two planets 
with the given mass ratios leads to  trapping into the $2:1$ 
commensurability. The resonance angles both undergo stable periodic 
librations as per section (\ref{mod})
 and the resonance is robust for 
the length of the 
simulation. We can conclude that  stable  resonance trapping  
would be  the probable outcome of the evolution
of  such a system. The finding of the system 
GJ876 near to or 
in such a resonance is probably simply a consequence of past tidally 
induced migration of the two planets into such a state. The magnitudes of 
the  eccentricities are in broad agreement with fitted model parameters to 
the  observations of GJ876 with $e_{1} \approx 0.1$ and $e_{2} \approx 
0.27$ (Marcy 
et al. 
\cite{Marcy4}).  The eccentricity ratio $e_{2} / e_{1} \approx 5.7.$ 
We comment that equation (\ref{eccpop}) which applies
in the small eccentricity limit  when one of $\phi, \psi$ is zero and the other $\pi$
gives $e_1/e_2 \sim 1/11.$
However, the fact that both angles librate about zero indicates,
within the context of the simple model,  that non resonant
orbital precession   needs to be incorporated
 and we should use equation (\ref{eccp})
to make a comparison with the simulation.
This is 
\begin{eqnarray}
& & e_1 e_2 a_1 M_* \omega_{pr1}+
 m_2 a_1  e_2 n_1 B \cos \phi = \nonumber \\ & & - m_1 a_2 e_1n_2 C\cos\psi
+e_1 e_2 a_1 M_* \omega_{pr2}. \end{eqnarray}
Since the librations of the angles are about a value close to  zero, we  replace them by zero
to make simple estimates.
The largest non central mass in the system is $m_1$ so we include
its effect  in causing non resonant orbital precession  for $m_2$
by  including a non zero $\omega_{pr2}.$
Also we include the corresponding effect on $m_1$ due to $m_2$
which produces a non zero value of $\omega_{pr1}.$
Then we have
  
\be \omega_{pr1}+  { m_2  n_1 B\over e_1  M_*}  = \omega_{pr2}- {n_2 m_1a_2 C\over e_2 a_1 M_*}.\ee
This condition equates the precession rate of the orbit of  $m_2$
on the right with that of $m_1$ on the left. In the former case the first contribution
comes from the time averaged orbit and the second is the resonant contribution.
In fact with $e_2 \sim 0.34,$ the orbits approach each other quite closely to within
$22$ percent of the  outer semi major axis.
But the resonant configuration avoids close encounters  keeping
the planets apart.  Thus we expect the resonant and non resonant contributions to the
precession rate to   show significant cancelation. The large  value of $e_2$ probably makes strict comparison
innacurate.  Nonetheless the simulation gives a precession  period of about $70$
orbits in a retrograde sense.  We can estimate that this  to be 
of the same magnitude as would be 
predicted from the second term on the left hand side which 
however  gives prograde precession.
Thus the non resonant
 precession rate due to $m_2$  needs to be
be comparable to  the resonant effect but of opposite sign.

Considering the precession rate of $m_2,$ we estimate that $\omega_{pr2}$ corresponds to
$70$ orbits in the prograde sense
while the resonant term corresponds to $70/2$ orbits in the retrograde sense. Thus the combined effect
produces a precession period of $70$ orbits in the retrograde
sense as seen in the simulation. 

  In summary our simulation gives plasuble 
eccentricity values for the two planets,
 that can be understood in outline   by use of a simplified  analytic 
theory and  are consistent with  the current observations.

\section{Orbit integrations}\label{Nb}  

In addition to the analytic model presented in section~(\ref{mod}) and the
hydrodynamic simulations presented in section~(\ref{simulation}),
we have also performed
three-body orbit integrations using a fifth-order Runge-Kutta scheme
(e.g. Press et al. 1993). 

The basic assumptions of the model are that the two planets exist
within the inner cavity of a tidally truncated disc that lies
exterior to the outer planet.
Tidal interaction with this disc causes inwards migration of the outer planet,
and also leads to eccentricity damping of the outer planet.
It is further assumed that as the planets migrate inwards and 
approach their final semi-major
axes, the disc disperses on some prescribed time scale $t_{disp}.$
In our
numerical calculations, a torque was applied
to the outermost planet such that it migrated inwards on a time scale
of $t_{mig}$ local orbital periods as defined in section(\ref{mod}), and a damping force was applied in
the radial direction to damp the eccentricity on a time scale of
$t_c$ local orbital periods also as defined in section (\ref{mod}). 

\begin{table}
\begin{center}
\begin{tabular}{r| r| r| r| r| r|  l}
Run & $ t_{mig}$ & $t_c $ & $ t_{disp}$ & $e_1$ & $e_2$ & $e_2/e_1$ 
  \\
 \hline
 R1  & $3.33\times 10^3 \;\;\;$& 900& $8 \times 10^3 \;\;$ & 0.095 & 0.41 &  4.3 \\
 R2  & $ 3.33\times 10^3 \;\;\;$& 250& $8 \times 10^3 \;\;$ & 0.05 & 0.3 &  6 \\
 R3  & $ 3.33 \times 10^2 \;\;\;$& 900& $8 \times 10^2 \;\;$ & 0.34 & 0.72 &  2.1 \\
 R4  & $ 10^3$& 270& $2.4 \times 10^3$ & 0.095 & 0.41 &  4.3 \\
 R5  & $ 3.33\times 10^3 \;\;\;$& 900& $0.0 \;\;\;\;\;$ & 0.095 & 0.41 &  4.3 \\
 \end{tabular}
 \end{center}
 \caption{This table shows the parameters used for the orbital
 integration runs. The first column gives the run label, the second column the
 migration time scale, the third column the circularisation time scale, and
 the fourth column gives the disc dispersal time. The fifth, sixth and seventh columns give the final eccentricities for the outer and inner planets, and
 the ratio of these eccentricities.}
 \label{tab1}
 \end{table}

These integrations used initial conditions  corresponding to
the more massive, outermost planet  being located initially at 5 AU, with the
lighter inner-most planet located initially at 2.5 AU. 
The planet masses adopted for the orbit integrations are the same 
as the minimum masses
reported for the planets in the system around GJ876 by 
Marcy et al. (\cite{Marcy4}) (i.e. 1.87 M$_{J}$
and 0.56 M$_{J}$). The stellar mass is taken to be 0.32 M$_{\odot}$.
Whilst these calculations provide only
a crude approximation to the detailed physics of disc--companion
interactions, their simplicity allows us to perform many calculations, covering
a wide area of parameter space, and also to run for  much 
longer time scales than is possible for simulations
of the type described in section(\ref{simulation}).

\begin{figure}
\epsfig{file=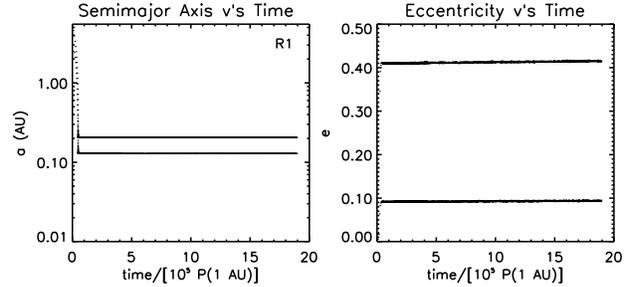,width=9cm}
\caption{This figure shows the evolution of the planet
semi-major axes and eccentricities for the run R1 shown in table~\ref{tab1}.}
\label{fig3}
\end{figure}

A number of calculations have been performed to examine the relationship
between the final values of $e_1$, $e_2$,  and their ratio
$e_1/e_2$,  to the various input parameters
$t_{mig}$, $t_c$, and $t_{disp}$.  The results of some of these calculations
are presented in table~\ref{tab1}, and are discussed below.
The unit of time used in the abscissa of the figures~\ref{fig3} to \ref{fig5}
is the orbital period for an object at 1 AU in orbit around a star with
mass $0.32$ M$_{\odot}$, and is denoted as P(1 AU).

\subsection{Dependence on migration and circularization  times}
Equation~\ref{e1} shows that the eccentricity
of the outer planet, $e_1$, depends on the ratio of $t_c/t_{mig}$. Here we
present results of simulations that explore how the eccentricity ratio
$e_2/e_1$ depends  on $t_c$ and $t_{mig}.$

\begin{figure}
\epsfig{file=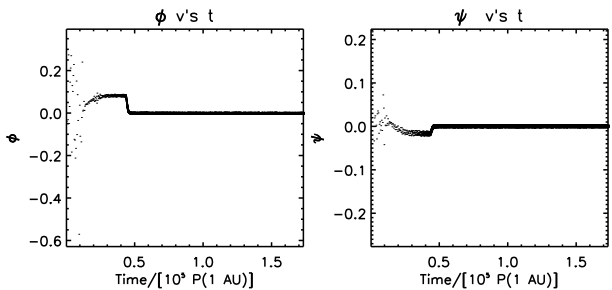,width=9cm}
\caption{This figure shows the evolution the resonant angles $\phi$ and $\psi$
for the run R1 described in table~\ref{tab1}. These show libration around
$\phi=\psi=0$, in agreement with the results from the hydrodynamic simulation
shown in figure~\ref{Fig1}.}
\label{fig4}
\end{figure}

\begin{figure}[h]
\epsfig{file=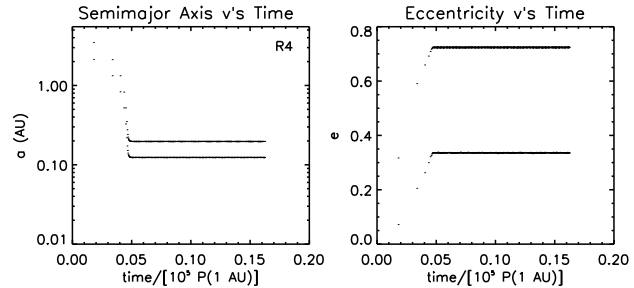,width=9cm}
\caption{This figure shows the evolution of the planet
semi-major axes and eccentricities for the run R3 shown in table~\ref{tab1}.}
\label{fig5}
\end{figure}

Figure~\ref{fig3} shows the evolution of the semi--major axes and
eccentricities for the run R1, whose model parameters are described in 
table~\ref{tab1}. This figure shows the inward migration of the outer
planet that subsequently locks to the inner planet as it reaches the 2:1
commensurability. The subsequent evolution is such that the two planets,
now resonantly locked, migrate inwards. The eccentricities result
from the balance between eccentricity driving through the resonant
interaction, and eccentricitiy damping due to the disc interaction.
As the planets approach their final semi-major axes, the effects of
migration and eccentricity damping are removed on a time scale of
$t_{disp}=8 \times 10^3$ local orbits, causing them to cease migration at
semi--major axes $a_1 \simeq 0.2$ and $a_2 \simeq 0.126$ which are values
appropriate to GJ876. The subsequent 
evolution beyond a time of $t \simeq 0.45$ P(1 AU) in figure~\ref{fig3} 
occurs in the absence of disc effects, and suggests a long--term
stability of the system given that it remains stable for $2 \times 10^7$ orbits
of the outer planet at its final semi-major axis.
Figure~\ref{fig4} shows the evolution of the
resonant angles $\phi$ and $\psi$, which librate about $\phi=\psi=0$ in 
agreement with the results presented in section~\ref{simulation} for the full
hydrodynamic simulations.

Figure~\ref{fig4} shows the results from run R2 described in table~\ref{tab1},
and illustrates the effect of reducing $t_{mig}$ while keeping $t_c$ constant.
As expected from equation~\ref{e1}, the eccentricity of the outer
planet increases from $e_1 \simeq 0.1$ to $e_1 \simeq 0.35$ when $t_{mig}$
changes from $ 3.33\times10^3$ to $3.33\times10^2$ local orbital periods. However, the ratio
$e_2/e_1$ has also changed from $e_2/e_1 \simeq 4$ to $e_2/e_1 \simeq 2$.

Calculation R3 illustrates the effect of keeping $t_{mig}$ constant
while changing $t_c$. As expected from equation~\ref{e1}, reducing $t_c$
leads to a reduction in $e_1$, since the disc model damps the eccentricity
more effectively. We also find that the ratio $e_2/e_1$ again changes,
it now being $e_2/e_1 \simeq 6$, as compared to $e_2/e_1 \simeq 4$ for run R1.

Equation~\ref{e1} predicts that keeping the ratio $t_{mig}/t_c$ constant,
but changing both $t_{mig}$ and $t_c$ independently, should leave $e_1$
unchanged. Calculation R4 indicates that this is what happens, and also
shows that the ratio $e_2/e_1$ remains unchanged.

Overall the results are entirely consistent with the analytic
predictions presented in section~\ref{mod} and with the hydrodynamical
simulations presented in section~\ref{simulation}.
Furthermore, they indicate that
the ratio $e_2/e_1$ scales rather weakly with $t_{mig}/t_c$.
We comment that from section~\ref{mod} equation~\ref{eccp} we expect the
eccentricity ratio to reach $\sim 11$ as $e_1 \rightarrow 0$.
Long--term stabilty of two--planet systems that become locked due to
disc--induced orbital migration is also indicated by our calculations.
In particlular run R1 covers
a time scale corresponding to $2 \times 10^7$ 
orbits of the outer planet in its
final configuration.
We find that it is possible to arrange $e_1$
to match the observed value of the outer planet in the GJ876 system
by fixing $t_{mig}$ and
choosing $t_c$ appropriately, but it is difficult to then obtain
a value for $e_2$ that matches the reported value of $e_2=0.27$.

\subsection{Dependence on Disc Dispersal Time Scale}
We have performed simulations to examine the effect of removing the
disc on different time scales on the stability
of the system. We find that the stability is largely
unaffected by the rate at which
the disc is removed. Figure~\ref{fig3} shows the evolution of the semi-major
axes and eccentricities from a run in which the disc was removed on a time scale
of $t_{disp}=8 \times 10^3$ local orbital periods.
In this case the disc dispersal
was switched on once the outer planet semi-major axis was $a_1 < 0.5$ AU.
The final semi-major axis of the planet is determined by the radius
at which the disc dispersal is initiated, and the time scale over which the
disc is removed. Calculation R5 was similar to R1 except that the disc
was removed instantaneously. The results presented in table~\ref{tab1}
show that this has little effect on the final outcome. A
slight increase in the scatter of the temporal distribution of
eccentricities was observed.  

\section{Summary and Discussion} \label{blah}
In this paper we have
considered two protoplanets gravitationally interacting with each
other and a protoplanetary disc. The two  planets orbit
interior to a tidally maintained disc cavity
while the disc interaction induces
inward migration. 

We have supposed that the   previous evolution of the system
 results in the the planets  getting  into a  configuration
with an orbital
separation just exceeding
that required for a $2:1$ commensurability.
This evolution is likely to have involved both accretion and migration.
Given the set up we consider a natural evolution is towards both
planets orbiting in a cavity outside of which orbits the protoplanetary disc.
Tidal interaction results in the inner disc material either being
expelled into the outer disc or accreting onto the central star.
Subsequently the outer planet migrates towards the inner one
as a result of interacting tidally  with the exterior disc.

 When the migration is slow enough,   we found that the
outer protoplanet  approached and became
locked into  a $2:1$ commensurability  with the inner one. 
This was maintained
in subsequent evolution.
We studied  the   nature of these interactions
using a simple analytic model,
hydrodynamic 2D simulations 
and  longer time  N body integrations. 
These all gave consistent results.

The magnitude of the stabilized  eccentricities 
was found to be  determined
by the ratio of the  migration rate to the circularization rate
induced in the outer planet orbit by the external disc.
The eccentricity ratio $e_1/e_2$ was found to vary with the magnitude of 
the ratio
being
more extreme for smaller eccentricities.
 
We  have applied our results to the recently discovered resonant planets
around GJ876.  Simulation shows that a disc with parameters expected for
protoplanetary discs causes trapping in the  $2:1$ commensurability when
the planets orbit in an inner cavity  and
that eccentricities  in the observed range
may be obtained. 
In this case the orbits were found to be aligned
with both resonant angles librating about zero.
The whole system then precessed in a retrograde sense.
Finally when the disc is removed
on a range of timescales the orbital configuration
has been found to be stable for up to $2 \times 10^7$ orbits subsequently.

\begin{acknowledgements} This work was supported by
PPARC grant number PPARC GR/L 39094.
We thank Udo Ziegler for making a FORTRAN Version of his code NIRVANA
publicly available.
\end{acknowledgements}

\end{document}